\documentclass[useAMS,usenatbib]{mn2e}

\RequirePackage[dvips]{graphicx}
\usepackage{color}
\usepackage{Times}

\def\apj{\emph{ApJ.}}
\def\aap{\emph{A.\& A.}}
\def\apjs{\emph{ApJS}}
\def\apss{\emph{Ap\&SS}}
\def\apjl{\emph{ApJ. Lett.}}

\def\mnras{\emph{MNRAS}}
\def\nat{\emph{Nature}}

\def\pasj{\emph{PASJ}}

\title[Lag--Luminosity Relation]{The Lag--Luminosity Relation in the GRB Source--Frame: \\An Investigation
with $Swift$ BAT Bursts}

\author[Ukwatta et al.]{T. N. Ukwatta$^{1,2}$\thanks{E-mail: tilan.ukwatta@gmail.com
(AVR)}, K. S. Dhuga$^{3}$, M. Stamatikos$^{2,4}$, C. D.
Dermer$^{5}$, T. Sakamoto$^{2,6}$,
\newauthor
E. Sonbas$^{2,6,7}$, W. C. Parke$^{3}$, L. C. Maximon$^{3}$, J. T.
Linnemann$^{1}$, P. N. Bhat$^{9}$,
\newauthor
A. Eskandarian$^{3}$, N. Gehrels$^{2}$, U. Abeysekara$^{1}$, K. Tollefson$^{1}$, and J. P. Norris$^{10}$\\
$^{1}$Department of Physics and Astronomy, Michigan State University, East Lansing, MI 48824, USA.\\
$^{2}$NASA Goddard Space Flight Center, Greenbelt, MD 20771, USA.\\
$^{3}$Department of Physics, The George Washington University, Washington, D.C. 20052, USA.\\
$^{4}$Center for Cosmology and Astro-Particle Physics (CCAPP)
Fellow, Department
of Physics, The Ohio State University, Columbus, OH 43210, USA.\\
$^{5}$Space Science Division, Code 7653, Naval Research
Laboratory, Washington, DC 20375, USA.\\
$^{6}$Center for Research and Exploration in Space Science and
Technology (CRESST), NASA Goddard Space Flight Center, Greenbelt,
MD 20771, USA.\\
$^{7}$University of Ad{\i}yaman, Department of Physics, 02040
Ad{\i}yaman, Turkey. \\
$^{8}$Universities Space Research Association, 10211 Wincopin
Circle, Suite 500, Columbia, MD 21044-3432, USA.\\
$^{9}$University of Alabama in Huntsville Center for Space Plasma
and Aeronomic Research, 320 Sparkman Dr. Huntsville AL, 35805,
USA.\\
$^{10}$Boise State University, Physics Department, Boise, ID 83725, USA.\\
}

\begin{document}



\maketitle

\label{firstpage}

\begin{abstract}
Spectral lag, which is defined as the difference in time of
arrival of high and low energy photons, is a common feature in
Gamma-ray Bursts (GRBs). Previous investigations have shown a
correlation between this lag and the isotropic peak luminosity for
long duration bursts. However, most of the previous investigations
used lags extracted in the observer-frame only. In this work
(based on a sample of 43 $Swift$ long GRBs with known redshifts),
we present an analysis of the lag-luminosity relation in the GRB
source-frame. Our analysis indicates a higher degree of
correlation $-0.82 \pm 0.05$ (chance probability of $\sim 5.5
\times 10^{-5}$) between the spectral lag and the isotropic peak
luminosity, $L_{\rm iso}$, with a best-fit power-law index of
$-1.2 \pm 0.2$, such that $L_{\rm iso}\propto {\rm lag}^{-1.2}$.
In addition, there is an anti-correlation between the source-frame
spectral lag and the source-frame peak energy of the burst
spectrum, $E_{\rm pk}(1+z)$.
\end{abstract}


\section{Introduction}
Gamma-ray Bursts (GRBs) are extremely energetic events and produce
highly diverse light curves. A number of empirical correlations
between various properties of the light curves and GRB energetics
have been discovered. However, the underlying physics of these
correlations is far from being understood.

One such correlation is the relation between isotropic peak
luminosity of long bursts and their spectral lags
\citep{norris2000}. Various authors have studied this relation
using arbitrary observer-frame energy-bands of various
instruments~\citep{Ukwatta2009lag, Hakkila2008, Schaefer2007,
gehrels2006, norris2002}. These investigations support the
existence of the relation, however with considerable scatter in
the extracted results. Recently, \cite{Margutti2010} investigated
spectral lags of X-ray flares and found that X-ray flares of long
GRBs also exhibit the lag-luminosity correlation observed in the
prompt emission.

The spectral lag is defined as the difference in time of arrival
of high and low energy photons and is considered to be positive
when the high-energy photons arrive earlier than the low energy
ones. Typically the spectral lag is extracted between two
arbitrary energy bands in the observer frame. However, because of
the redshift dependence of GRBs, these two energy bands can
correspond to a different pair of energy bands in the GRB
source-frame, thus potentially introducing an arbitrary energy
dependence to the extracted spectral lag.

In order to explore whether the lag-luminosity relation is
intrinsic to the GRB, it is preferable to extract spectral lags in
the source frame as opposed to the observer frame. At least two
corrections are needed to accomplish this: 1) Correct for the time
dilation effect (z-correction), and 2) Take into account the fact
that for GRBs with various redshifts, observed energy bands
correspond to different energy bands at the GRB source-frame
(K-correction; \cite{gehrels2006}).

The first correction is straightforward and is achieved by
multiplying the extracted lag value (in the observer-frame) by
$(1+z)^{-1}$. The second correction, on the other hand, is not so
straightforward. \cite{gehrels2006} attempted to approximately
correct the spectral lag by multiplying the lag value (in the
observer-frame) by $(1+z)^{0.33}$. We note here that this
correction is based on the assumption that the spectral lag is
proportional to the pulse width and that the pulse width itself is
proportional to the energy ~\citep{Zhang2009, Fenimore1995}. These
approximations depend on clearly identifying corresponding pulses
in the light curves of each energy band, and may be of limited
validity for a large fraction of GRBs in which the light curves
are dominated by overlapping multi-pulse structures.

Using a sample of 31 $Swift$ GRBs, \cite{Ukwatta2009lag}
(hereafter U10) found that the correlation coefficient improves
significantly after the z-correction is applied. However, this
correlation does not improve further after the application of the
K-correction as defined by \cite{gehrels2006}.

An alternative is to make the K-correction by choosing two
appropriate energy bands fixed in the GRB source-frame and
projecting these bands into the observer-frame using the relation
$E_{\rm observer}=E_{\rm source}/(1+z)$. \cite{Ukwatta2010lag}
used this method for the first time to investigate the
lag-luminosity relation in the source-frame of the GRB. They
selected two source-frame energy bands (100 -- 200 keV and 300 --
400 keV) and used background subtracted as well as non-background
subtracted $Swift$ data to extract lags. Non-background subtracted
data were used to improve the signal-to-noise ratio for weak
bursts. They found that the source-frame relation seems a bit
tighter, but with a slope consistent with previous studies.
\cite{Arimoto2010} also looked at a limited sample of HETE-II
bursts (8 GRBs) both in the observer-frame and the source-frame
and concluded that there is no significant effect from the
redshift. However, the redshift distribution of their burst sample
is very narrow and peaks around one. In contrast to
\cite{Ukwatta2010lag}, in this study we used only background
subtracted data and measured the lag between source-frame energy
bands 100 -- 150 keV and 200 -- 250 keV (the reason for selecting
these particular energy bands is described in
$\S$~\ref{methodology}) for a sample of 43 $Swift$ bursts with
spectroscopic redshifts.

In this work we have investigated only long GRBs, i.e., bursts
with duration greater than $\sim 2$ seconds. It is rather
difficult to test the Lag-$L_{\rm iso}$ relation effectively for
short GRBs due to a lack of spectroscopically measured redshifts.
None of the short bursts detected so far have any redshift
measurements obtained from a spectroscopic analysis of their
optical afterglow. Moreover, it has been shown that short GRBs
have either small or negligible lags~\citep{norris2006,
Zhang2006}. According to the Lag-$L_{\rm iso}$ relation, these
small lag values imply that short bursts to be highly luminous.
However, based on the redshift measurements of their host galaxies
we can show that short GRBs are generally less luminous than long
bursts. Hence short bursts seem to not follow the lag-luminosity
relation (\cite{gehrels2006}).

The structure of this paper is the following: In
$\S$~\ref{methodology} we discuss briefly our methodology for
extracting spectral lags. In $\S$~\ref{results} we present our
results for a sample of 43 $Swift$ GRBs. We discuss our results
with two candidate models in $\S$~\ref{discussion}. Finally, in
the last section ($\S$~\ref{conclusion}) we summarize our results
and conclusions. Throughout this paper, the quoted uncertainties
are at the 68\% confidence level.

\section[]{Methodology} \label{methodology}

\begin{figure*}
\centering
\includegraphics[width=140mm]{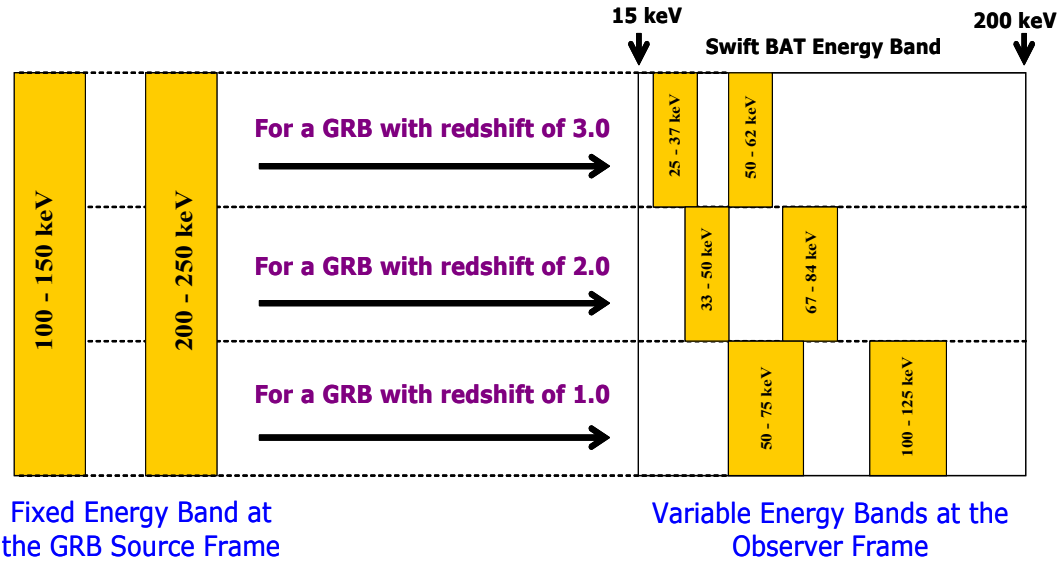}
\caption{Fixed energy bands at the GRB source-frame are projected
to various energy bands at the observer-frame, depending on the
redshift.}\label{fig01}
\end{figure*}

The $Swift$ Burst Alert Telescope (BAT) is a highly sensitive
instrument using a coded-mask aperture~\citep{barthelmy2005}. BAT
uses the shadow pattern resulting from the coded mask to
facilitate localization of the source. When a gamma-ray source
illuminates the coded mask, it casts a shadow onto a
position-sensitive detector. The shadow cast depends on the
position of the gamma-ray source on the sky. If one knows the tile
pattern in the coded mask and the geometry of the detector, it is
possible to calculate the shadow patterns created by all possible
points in the sky using a ray-tracing algorithm. Hence, by
correlating the observed shadow, with the pre-calculated shadow,
one can find the location of the source. However, each detector
can be illuminated by many sources and a given source can
illuminate many detectors. Hence, in order to disentangle each sky
position, special algorithms have been developed and integrated in
to the data analysis software by the $Swift$ BAT team.

To generate background-subtracted light curves we used a process
called mask weighting. The mask weighting assigns a ray-traced
shadow value for each individual event, which then enables the
user to calculate light curves or spectra. We used the
\texttt{batmaskwtevt} and \texttt{batbinevt} tasks in FTOOLS to
generate mask weighted, background-subtracted light curves, for
various observer-frame energy bands, as shown in
Table~\ref{tab:ebands}. These are the energy bands that correspond
to fixed energy bands in the source-frame i.e. $100-150$ and
$200-250$ keV. These particular energy bands were selected so that
after transforming to the observer-frame they lie in the
detectable energy range of the $Swift$ BAT instrument (see
Fig.~\ref{fig01}). Even though the BAT can detect photons up to
350 keV, we limited the upper-boundary to 200 keV in the
observer-frame. This is because the mask-weighted effective area
of the detector falls rapidly after 200 keV and as a result the
contribution to the light curve from energies greater than
$\sim$200 keV (in observer-frame) is
negligible~\citep{Sakamoto2011}.

\begin{table*}
\centering
\begin{minipage}{120mm}
  \caption{The observer-frame energy bands and energy gaps
(the energy difference between the mid-points of energy bands) for
bursts in the sample.\label{tab:ebands}}
  \begin{tabular}{@{}llccc@{}}
  \hline
  \hline
GRB & Redshift & Low Energy Band (keV) & High Energy Band (keV) &
Energy Gap (keV)\\
 \hline \hline
GRB050401 &  $ 2.899^{1}$ & 26-38 & 51-64 & 26 \\
GRB050603 &  $ 2.821^{2}$ & 26-39 & 52-65 & 26 \\
GRB050922C &  $ 2.199^{3}$ & 31-47 & 63-78 & 32 \\
GRB051111 &  $ 1.549^{4}$ & 39-59 & 78-98 & 39 \\
GRB060206 &  $ 4.056^{5}$ & 20-30 & 40-49 & 20 \\
GRB060210 &  $ 3.913^{6}$ & 20-31 & 41-51 & 21 \\
GRB060418 &  $ 1.490^{7}$ & 40-60 & 80-100 & 40 \\
GRB060904B &  $ 0.703^{8}$ & 59-88 & 117-147 & 59 \\
GRB060908 &  $ 1.884^{9}$ & 35-52 & 69-87 & 35 \\
GRB060927 &  $ 5.464^{10}$ & 15-23 & 31-39 & 16 \\
GRB061007 &  $ 1.262^{11}$ & 44-66 & 88-111 & 45 \\
GRB061021 &  $ 0.346^{12}$ & 74-111 & 149-186 & 75 \\
GRB061121 &  $ 1.315^{13}$ & 43-65 & 86-108 & 43 \\
GRB070306 &  $ 1.496^{14}$ & 40-60 & 80-100 & 40 \\
GRB071010B &  $ 0.947^{15}$ & 51-77 & 103-128 & 52 \\
GRB071020 &  $ 2.145^{16}$ & 32-48 & 64-79 & 32 \\
GRB080319B &  $ 0.937^{17}$ & 52-77 & 103-129 & 52 \\
GRB080319C &  $ 1.949^{18}$ & 34-51 & 68-85 & 34 \\
GRB080411 &  $ 1.030^{19}$ & 49-74 & 99-123 & 50 \\
GRB080413A &  $ 2.433^{20}$ & 29-44 & 58-73 & 29 \\
GRB080413B &  $ 1.101^{21}$ & 48-71 & 95-119 & 48 \\
GRB080430 &  $ 0.767^{22}$ & 57-85 & 113-141 & 56 \\
GRB080603B &  $ 2.689^{23}$ & 27-41 & 54-68 & 27 \\
GRB080605 &  $ 1.640^{24}$ & 38-57 & 76-95 & 38 \\
GRB080607 &  $ 3.036^{25}$ & 25-37 & 50-62 & 25 \\
GRB080721 &  $ 2.591^{26}$ & 28-42 & 56-70 & 28 \\
GRB080916A &  $ 0.689^{27}$ & 59-89 & 118-148 & 59 \\
GRB081222 &  $ 2.770^{28}$ & 27-40 & 53-66 & 26 \\
GRB090424 &  $ 0.544^{29}$ & 65-97 & 130-162 & 65 \\
GRB090618 &  $ 0.540^{30}$ & 65-97 & 130-162 & 65 \\
GRB090715B &  $ 3.000^{31}$ & 25-38 & 50-63 & 25 \\
GRB090812 &  $ 2.452^{32}$ & 29-43 & 58-72 & 29 \\
GRB090926B &  $ 1.240^{33}$ & 45-67 & 89-112 & 45 \\
GRB091018 &  $ 0.971^{34}$ & 51-76 & 101-127 & 51 \\
GRB091020 &  $ 1.710^{35}$ & 37-55 & 74-92 & 37 \\
GRB091024 &  $ 1.091^{36}$ & 48-72 & 96-120 & 48 \\
GRB091029 &  $ 2.752^{37}$ & 27-40 & 53-67 & 27 \\
GRB091208B &  $ 1.063^{38}$ & 48-73 & 97-121 & 49 \\
GRB100621A &  $ 0.542^{39}$ & 65-97 & 130-162 & 65 \\
GRB100814A &  $ 1.440^{40}$ & 41-61 & 82-102 & 41 \\
GRB100816A &  $ 0.800^{41}$ & 56-83 & 111-139 & 56 \\
GRB100906A &  $ 1.727^{42}$ & 37-55 & 73-92 & 37 \\
GRB110205A &  $ 2.220^{43}$ & 31-47 & 62-78 & 31 \\
GRB110213A &  $ 1.460^{44}$ & 41-61 & 81-102 & 41 \\
\hline
\end{tabular}
\\{(1) \cite{Watson2006}; (2) \cite{2005GCN..3520....1B}; (3)
\cite{Piranomonte2008}; (4) \cite{Penprase2006}; (5)
\cite{Fynbo2009}; (6) \cite{Fynbo2009}; (7)
\cite{2006GCN..5002....1P}; (8) \cite{Fynbo2009}; (9)
\cite{Fynbo2009}; (10) \cite{Fynbo2009}; (11) \cite{Fynbo2009};
(12) \cite{Fynbo2009}; (13) \cite{Fynbo2009}; (14)
\cite{Jaunsen2008}; (15) \cite{Cenko2007}; (16)
\cite{2007GCN..6952....1J}; (17) \cite{DElia2009}; (18)
\cite{Fynbo2009}; (19) \cite{Fynbo2009}; (20) \cite{Fynbo2009};
(21) \cite{Fynbo2009}; (22) \cite{2008GCN..7654....1C}; (23)
\cite{Fynbo2009}; (24) \cite{Fynbo2009}; (25)
\cite{Prochaska2009}; (26) \cite{Fynbo2009}; (27)
\cite{Fynbo2009}; (28) \cite{Cucchiara2008}; (29)
\cite{2009GCN..9243....1C}; (30) \cite{2009GCN..9518....1S}; (31)
\cite{2009GCN..9673....1S}; (32) \cite{2009GCN..9771....1D}; (33)
\cite{2009GCN..9947....1F}; (34) \cite{2009GCN.10038....1C}; (35)
\cite{2009GCN.10053....1X}; (36) \cite{2009GCN.10065....1C}; (37)
\cite{2009GCN.10100....1C}; (38) \cite{2009GCN.10263....1W}; (39)
\cite{2010GCN.10876....1M}; (40) \cite{2010GCN.11089....1O}; (41)
\cite{2010GCN.11230....1T}; (42) \cite{2011GCN.11638....1C}; (43)
\cite{2011GCN.11708....1M}. }
\end{minipage}
\end{table*}

The spectral lags were extracted using the improved
cross-correlation function (CCF) analysis method described in U10.
In this method, the spectral lag is defined as the time delay
corresponding to the global maximum of the cross-correlation
function. The CCF with a delay index $d$ is defined as,
\begin{equation}\label{eq:no1}
CCF(d, x, y)=\frac{\sum_{\small i=\textrm{max}(1, 1-d)}^{\small
\textrm{min}(N, N-d)}x_i \, y_{i+d}}{\sqrt{\sum_{i}x_i^2 \,
\sum_{i}y_{i}^2}}
\end{equation}
where $x_i$ and $y_i$ are two sets of time-sequenced data spread
over $N$ bins. The time delay is obtained by multiplying $d$ by
the time bin size of the light curves. A Gaussian curve was fitted
to the CCF (plotted as a function of time delay) to extract the
spectral lag. The uncertainty in the spectral lag is obtained by
simulating 1,000 light curves using the Monte Carlo technique (see
U10 for more details).

The isotropic peak luminosity ($L_{\rm iso}$) and its uncertainty
for each GRB is obtained using the method described in U10. In
essence, a typical GRB spectrum can be described by the Band
function~\citep{band1993}, for the photon flux per unit photon
energy using
\begin{equation}
N(E) = \left\{ \begin{array}{ll}A
(\frac{E}{100\,\rm{keV}})^{\,\alpha}\,e^{-(2+\alpha)E/E_{\rm
pk}},\,\small E\,\le\,\big(\frac{\alpha -
\beta}{2+\alpha}\big)E_{\rm pk}& \\
A (\frac{E}{100\,\rm{keV}})^{\,\beta}\,[\frac{(\alpha-\beta)E_{\rm
pk}}{(2+\alpha)100\,\rm{keV}}]^{\alpha-\beta}\,e^{(\beta-\alpha)},
\,\rm else, &
\end{array} \right.
\end{equation}
which has four model parameters: the amplitude (A), the low-energy
spectral index ($\alpha$), the high-energy spectral index
($\beta$) and the peak ($E_{\rm pk}$) of $E^2 N(E)$ spectrum (also
called the $\nu F_{\nu}$ spectrum, apart from a factor of Planck's
constant). Using these spectral parameters the observed peak flux
can be calculated for the source-frame energy-range $E_{1}
=1.0\,\rm{keV}$ to $E_{2} = 10,000\,\rm{keV}$ using
\begin{equation}
f_{\rm obs} = \int_{E_{1}/(1+z)}^{E_{2}/(1+z)}\,N(E)E\, dE.
\end{equation}
The isotropic peak luminosity is defined by
\begin{equation}
L_{\rm iso}= 4 \pi d_L^{\,2} \, f_{\rm{obs}}
\end{equation}
where $d_L$ is the luminosity distance:
\begin{equation}
d_L=\frac{(1+z)c}{H_0}\int_{0}^{z} \frac{dz'}{\sqrt{\Omega_M
(1+z')^3 + \Omega_L}}.
\end{equation}
For the current universe we take $\Omega_M = 0.27$, $\Omega_L =
0.73$ and the Hubble constant $H_0$ to be
$70\,\rm(kms^{-1})/Mpc$~\citep{Komatsu2009}. For more details of
the $L_{\rm iso}$ calculation see U10.

\section[]{Results} \label{results}

We employed an additional 12 long bursts to the GRB sample (31
GRBs) that was used in U10, which increased the total sample to
43. This sample has redshifts ranging from 0.346 (GRB 061021) to
5.464 (GRB 060927) with an average redshift of $\sim$2.0. The
spectral information for the additional 12 bursts used in this
paper is given in Table~\ref{tab:spectral_info}. The calculated
peak isotropic luminosities, spanning three orders of magnitude,
are given in U10 and Table~\ref{tab:spectral_info}.

\begin{table*}
\centering
\begin{minipage}{170mm}
  \caption{GRB redshift and spectral information. Note that uncertainties of parameters that are
reported with 90\% confidence level have been reduced to
1-$\sigma$ level for consistency.\label{tab:spectral_info}}
  \begin{tabular}{@{}llccccc@{}}
  \hline
  \hline
GRB & Peak Flux $^{\rm a}$ & $E_{\rm pk}$$^{\,b}$ &
$\alpha$ & $\beta^{\,c}$ & $L_{\rm iso}$ erg/s & Reference \\
 \hline \hline
GRB090812 & $ 3.60 \pm 0.13 $ & $  572^{+ 99}_{- 156} $ & $-1.03^{+0.04}_{-0.04} $ & $-2.50^{+0.16}_{-0.16} $ & $(7.86^{+1.95}_{-0.87} ) \times 10^{ 52}$  & \cite{9821}; \cite{9775} \\
GRB090926B & $ 3.20 \pm 0.19 $ & $  91^{+ 1}_{- 1} $ & $-0.13^{+0.04}_{-0.04} $ & $-2.36^{+0.31}_{-0.31} $ & $(5.22^{+3.88}_{-0.82} ) \times 10^{ 51}$  & \cite{9957}; \cite{9939} \\
GRB091018 & $ 10.30 \pm 0.25 $ & $  28^{+ 10}_{- 6} $ & $-1.53^{+0.24}_{-0.37} $ & $-2.44^{+0.15}_{-0.15} $ & $(6.96^{+1.76}_{-0.58} ) \times 10^{ 51}$  & \cite{10045}; \cite{10040} \\
GRB091020 & $ 4.20 \pm 0.19 $ & $  47^{+ 4}_{- 4} $ & $-0.20^{+0.25}_{-0.25} $ & $-1.70^{+0.01}_{-0.01} $ & $(2.81^{+0.19}_{-0.16} ) \times 10^{ 52}$  & \cite{10095}; \cite{10051} \\
GRB091024 & $ 2.00 \pm 0.19 $ & $  500^{+ 100}_{- 100} $ & $-1.10^{+0.13}_{-0.13} $ & $-2.36^{+0.31}_{-0.31} $ & $(5.56^{+2.43}_{-0.89} ) \times 10^{ 51}$  & \cite{10083}; \cite{10072} \\
GRB091029 & $ 1.80 \pm 0.06 $ & $  61^{+ 10}_{- 10} $ & $-1.46^{+0.17}_{-0.17} $ & $-2.36^{+0.31}_{-0.31} $ & $(1.67^{+0.60}_{-0.15} ) \times 10^{ 52}$  &  \cite{10103} \\
GRB091208B & $ 15.20 \pm 0.63 $ & $  124^{+ 12}_{- 12} $ & $-1.44^{+0.04}_{-0.04} $ & $-2.32^{+0.29}_{-0.12} $ & $(1.68^{+0.65}_{-0.09} ) \times 10^{ 52}$  & \cite{10266}; \cite{10265} \\
GRB100621A & $ 12.80 \pm 0.19 $ & $  95^{+ 8}_{- 11} $ & $-1.70^{+0.08}_{-0.08} $ & $-2.45^{+1.44}_{-1.44} $ & $(2.55^{+0.83}_{-0.34} ) \times 10^{ 51}$  & \cite{10882}; \cite{291} \\
GRB100814A & $ 2.50 \pm 0.13 $ & $  106^{+ 7}_{- 8} $ & $-0.64^{+0.08}_{-0.09} $ & $-2.02^{+0.08}_{-0.06} $ & $(8.27^{+1.13}_{-0.69} ) \times 10^{ 51}$  & \cite{11099}; \cite{11094} \\
GRB100906A & $ 10.10 \pm 0.25 $ & $  180^{+ 25}_{- 28} $ & $-1.10^{+0.06}_{-0.06} $ & $-2.20^{+0.19}_{-0.13} $ & $(4.90^{+1.23}_{-0.43} ) \times 10^{ 52}$  & \cite{11251}; \cite{11233} \\
GRB110205A & $ 3.60 \pm 0.13 $ & $  222^{+ 46}_{- 46} $ & $-1.52^{+0.09}_{-0.09} $ & $-2.36^{+0.31}_{-0.31} $ & $(2.78^{+0.57}_{-0.20} ) \times 10^{ 52}$  & \cite{11659} \\
GRB110213A & $ 1.60 \pm 0.38 $ & $  98^{+ 4}_{- 5} $ & $-1.44^{+0.03}_{-0.03} $ & $-2.36^{+0.31}_{-0.31} $ & $(3.53^{+1.97}_{-0.53} ) \times 10^{ 51}$  &  \cite{11727}; \cite{11714} \\
\hline
\end{tabular}
\\
{ $^a \,$ 1-second peak photon flux measured in $\rm photons\,
\,cm^{-2} \, s^{-1}$ in the energy range $15-150$ keV.

$^b \,$ Peak energy, $E_{\rm pk}$, is given in keV.


$^c \,$ Values in brackets indicates estimated high-energy photon
index, $\beta$, which is the mean value of the BATSE $\beta$
distribution \citep{Kaneko2006,Sakamoto2009}. }
\end{minipage}
\end{table*}

By choosing appropriate energy bands in the observer-frame
(according to the redshift of each burst), we extracted
mask-weighted background-subtracted light curves for the selected
source-frame energy bands 100--150 and 200--250 keV. The
observer-frame energy bands used for each burst are shown in
Table~\ref{tab:ebands}. Note that the energy gap between the
mid-point of the two source-frame energy bands is fixed at 100 keV
whereas in the observer-frame, as expected, this gap varies
depending on the redshift of each burst (see the
Table~\ref{tab:ebands}). For example, in GRB 060927, this gap is
16 keV and in GRB 061021, it is 75 keV. This is in contrast to the
spectral lag extractions performed in the observer-frame where
this gap is treated as a constant.

The extracted spectral lags for the source-frame energy bands
100--150 and 200--250 keV are listed in Table~\ref{tab:lag}. The
$Swift$ BAT trigger ID, the segment of the light curve used for
the lag extraction ($T+X_{\rm S}$ and $T+X_{\rm E}$, $T$ is the
trigger time), the time binning of the light curve, and the
Gaussian curve fitting range of the CCF vs time delay plot (with
start time, and end time denoted as $LS$ and $LE$ respectively)
are also given in Table~\ref{tab:lag}. Of 43 bursts in the sample
there are 24 bursts which have lags greater than zero. The
remaining 19 bursts either have lags consistent with zero (16
bursts) or negative values (3 bursts).

For the 24 bursts which have positive lags with significance
1-$\sigma$ or greater (see Table~\ref{tab:lag}), we find that the
redshift corrected lag is anti-correlated with $L_{\rm iso}$. The
correlation coefficient for this relation is -0.82 $\pm$ 0.05 with
a chance probability of $\sim 5.54 \times 10^{-5}$. The extracted
correlation coefficient is significantly higher than the
correlation coefficient (averaged over the six combinations of
standard BAT energy channels) of $\sim \, -0.68$ reported in U10.
Various correlation coefficients of the relation are shown in
Table~\ref{corr_table01}, where uncertainties in the correlation
coefficients were obtained through a Monte Carlo simulation
utilizing uncertainties in $L_{\rm iso}$ and the lag values. The
null probability that the correlation occurs due to random chance
is also given for each coefficient type.

\begin{figure*}
\centering
\includegraphics[width=120mm]{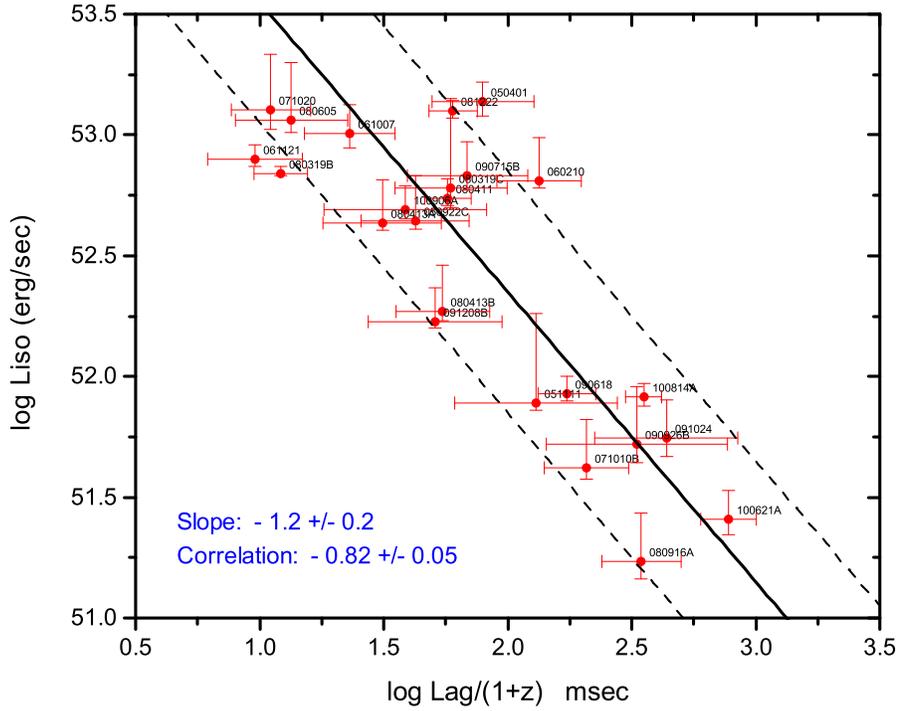}
\caption{The spectral lags between the source-frame energy range
bands $100-150$ keV and $200-250$ keV and the isotropic peak
luminosity are plotted in a log-log plot.}\label{fig02}
\end{figure*}


Fig.~\ref{fig02} shows a log-log plot of isotropic peak luminosity
vs redshift-corrected spectral lag. The solid line shows the
following best-fit power-law curve:

\begin{equation}\label{eq:no2}
\log \bigg(\frac{L_{\rm iso}}{{\rm erg/s}}\bigg) = (54.7 \pm 0.4)
- (1.2 \pm 0.2) \log \frac{Lag/({\rm ms})}{1+z}.
\end{equation}

Since there is considerable scatter, the uncertainties of the fit
parameters are multiplied by a factor of $\sqrt{\rm
\chi^2/ndf}=\sqrt{84.36/22}=1.96$. The dash lines indicate the
estimated 1-$\sigma$ confidence level, which is obtained from the
cumulative fraction of the residual distribution taken from 16\%
to 84\%.

The best-fit power-law index ($-1.2 \pm 0.2$) is consistent with
observer-frame results obtained by \cite{norris2000} ($\sim \,
-1.14$) and the average power-law index of $-1.4 \pm 0.3$ reported
in U10.

\begin{table*}
\centering
\begin{minipage}{150mm}
  \caption{Source-frame spectral lag values of long duration $Swift$ BAT GRBs
\label{tab:lag}}
  \begin{tabular}{@{}llllccccc@{}}
  \hline \hline
GRB & Trigger ID & $T+X_{\rm S}$ (s) & $T+X_{\rm E}$ (s) & Bin
Size (ms) & LS (s) & LE (s) & Lag Value (ms) & Significance \\
 \hline \hline

GRB050401 & 113120 & 23.03 & 29.43 & 64 & -2.00 & 2.00 & 310$\pm$145 & 2.14\\
GRB050603 & 131560 & -3.83 & 3.08 & 16 & -0.40 & 0.40 & -16$\pm$21 & -0.76\\
GRB050922C & 156467 & -2.70 & 2.94 & 16 & -1.00 & 1.00 & 136$\pm$68 & 2.00\\
GRB051111 & 163438 & -6.96 & 28.62 & 64 & -4.00 & 4.00 & 333$\pm$251 & 1.33\\
GRB060206 & 180455 & -1.29 & 8.18 & 16 & -2.00 & 2.00 & 86$\pm$111 & 0.77\\
GRB060210 & 180977 & -3.37 & 5.08 & 128 & -4.00 & 4.00 & 658$\pm$259 & 2.54\\
GRB060418 & 205851 & -7.66 & 33.04 & 64 & -2.00 & 2.00 & -110$\pm$106 & -1.04\\
GRB060904B & 228006 & -1.97 & 10.32 & 512 & -6.00 & 6.00 & 124$\pm$436 & 0.28\\
GRB060908 & 228581 & -10.91 & 3.68 & 32 & -2.00 & 2.00 & 78$\pm$124 & 0.63\\
GRB060927 & 231362 & -1.69 & 8.04 & 32 & -1.00 & 1.00 & 18$\pm$75 & 0.24\\
GRB061007 & 232683 & 23.86 & 65.08 & 4 & -0.20 & 0.20 & 52$\pm$22 & 2.36\\
GRB061021 & 234905 & -0.46 & 14.64 & 512 & -4.00 & 4.00 & -430$\pm$975 & -0.44\\
GRB061121 & 239899 & 60.44 & 80.66 & 4 & -0.20 & 0.20 & 22$\pm$10 & 2.20\\
GRB070306 & 263361 & 90.00 & 118.42 & 32 & -4.00 & 2.00 & -362$\pm$247 & -1.47\\
GRB071010B & 293795 & -1.70 & 17.24 & 64 & -2.00 & 2.00 & 404$\pm$159 & 2.54\\
GRB071020 & 294835 & -3.22 & 1.14 & 4 & -0.20 & 0.40 & 35$\pm$13 & 2.69\\
GRB080319B & 306757 & -2.85 & 57.57 & 4 & -0.10 & 0.14 & 23$\pm$6 & 3.83\\
GRB080319C & 306778 & -0.77 & 13.31 & 32 & -1.00 & 1.00 & 174$\pm$91 & 1.91\\
GRB080411 & 309010 & 38.46 & 48.45 & 4 & -0.50 & 0.50 & 116$\pm$25 & 4.64\\
GRB080413A & 309096 & -0.42 & 9.05 & 8 & -1.00 & 1.00 & 107$\pm$59 & 1.81\\
GRB080413B & 309111 & -1.44 & 4.96 & 32 & -1.00 & 1.00 & 115$\pm$50 & 2.30\\
GRB080430 & 310613 & -1.24 & 12.84 & 256 & -4.00 & 4.00 & 91$\pm$431 & 0.21\\
GRB080603B & 313087 & -0.54 & 5.10 & 16 & -1.00 & 1.00 & 5$\pm$59 & 0.08\\
GRB080605 & 313299 & -5.46 & 15.53 & 8 & -0.20 & 0.20 & 35$\pm$18 & 1.94\\
GRB080607 & 313417 & -6.13 & 12.05 & 8 & -0.50 & 0.50 & 26$\pm$30 & 0.87\\
GRB080721 & 317508 & -3.39 & 8.64 & 64 & -2.00 & 2.00 & -86$\pm$110 & -0.78\\
GRB080916A & 324895 & -2.66 & 39.58 & 128 & -2.00 & 4.00 & 585$\pm$214 & 2.73\\
GRB081222 & 337914 & -0.80 & 15.58 & 4 & -1.00 & 1.00 & 227$\pm$51 & 4.45\\
GRB090424 & 350311 & -0.94 & 4.95 & 16 & -0.20 & 0.20 & 14$\pm$14 & 1.00\\
GRB090618 & 355083 & 46.01 & 135.35 & 8 & -2.00 & 2.00 & 267$\pm$72 & 3.71\\
GRB090715B & 357512 & -4.80 & 21.06 & 16 & -2.00 & 3.00 & 275$\pm$155 & 1.77\\
GRB090812 & 359711 & -6.93 & 41.20 & 256 & -6.00 & 6.00 & -22$\pm$202 & -0.11\\
GRB090926B & 370791 & -22.00 & 36.00 & 512 & -10.00 & 8.00 & 746$\pm$627 & 1.19\\
GRB091018 & 373172 & -0.28 & 2.92 & 64 & -2.00 & 1.00 & 143$\pm$297 & 0.48\\
GRB091020 & 373458 & -2.54 & 13.84 & 128 & -3.00 & 2.00 & -187$\pm$177 & -1.06\\
GRB091024 & 373674 & -9.58 & 27.29 & 512 & -10.00 & 10.00 & 912$\pm$604 & 1.51\\
GRB091029 & 374210 & -4.03 & 38.98 & 256 & -10.00 & 10.00 & -112$\pm$395 & -0.28\\
GRB091208B & 378559 & 7.66 & 10.61 & 64 & -1.00 & 1.00 & 105$\pm$66 & 1.59\\
GRB100621A & 425151 & -6.79 & 40.31 & 256 & -3.00 & 3.00 & 1199$\pm$311 & 3.86\\
GRB100814A & 431605 & -4.40 & 29.39 & 256 & -4.00 & 4.00 & 862$\pm$147 & 5.86\\
GRB100906A & 433509 & -1.49 & 26.16 & 128 & -2.00 & 2.00 & 105$\pm$79 & 1.33\\
GRB110205A & 444643 & 118.89 & 293.99 & 64 & -1.00 & 1.00 & -29$\pm$52 & -0.56\\
GRB110213A & 445414 & -3.42 & 5.29 & 512 & -3.00 & 3.50 & 602$\pm$746 & 0.81\\

\hline
\end{tabular}
\end{minipage}
\end{table*}

\begin{table}
\centering
 \begin{minipage}{84mm}
  \caption{Correlation coefficients of the lag--luminosity relation.
  \label{corr_table01}}
  \begin{tabular}{@{}lll@{}}
  \hline
  Coefficient Type     &  Correlation Coefficient & Null Probability \\
 \hline
Pearson's $r$          & - 0.82$\pm$0.05 & $5.54\,\times\,10^{-5}$\\
Spearman's $r_{\rm s}$ & - 0.70$\pm$0.06 & $1.49\,\times\,10^{-4}$\\
Kendall's $\tau$       & - 0.50$\pm$0.05 & $6.63\,\times\,10^{-4}$\\
\hline
\end{tabular}
\end{minipage}
\end{table}

\section[]{Discussion}\label{discussion}

\subsection{Spectral Lags: Observer-frame versus Source-frame}

\begin{figure*}
\centering
\includegraphics[width=120mm]{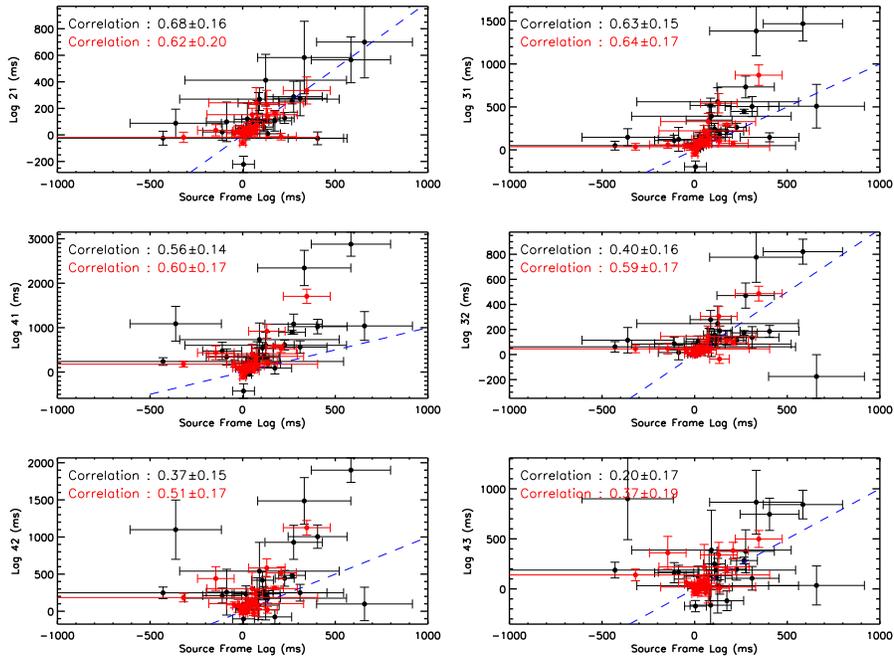}
\caption{All combinations of fixed observer-frame energy channel
(canonical BAT energy bands: channel 1 (15--25 keV), 2 (25--50
keV), 3 (50--100 keV) and 4 (100--200 keV)) spectral lag values as
a function of fixed source-frame energy channel (between 100--150
keV and 200--250 keV) lag values. Black and red data points and
labels corresponds to redshift uncorrected and corrected cases
respectively. The blue dashed line corresponds to the equality
line of the two parameters in each panel.}\label{Lag_Obs_Src_31}
\end{figure*}

U10 extracted spectral lags in fixed energy bands in the
observer-frame and in this work for the same sample of 31 bursts
we extracted lags in fixed energy bands in the source-frame. In
the observer-frame case, there are four energy channels (canonical
BAT energy bands: channel 1 (15--25 keV), 2 (25--50 keV), 3
(50--100 keV) and 4 (100--200 keV)), thus six lag extractions per
burst. It is interesting to study to what degree these different
lags correlate with source-frame lags (between fixed source-frame
energy channels 100--150 keV and 200--250 keV). In
Fig.~\ref{Lag_Obs_Src_31} we show all combinations of
observer-frame lags as a function of source-frame lags. The red
data points show lags with the time-dilation correction due to
cosmological redshift and black data points show lags without the
time-dilation correction. From Fig.~\ref{Lag_Obs_Src_31} it is
clear that all plots show some correlation both in the
time-dilation corrected (shown in red) and time-dilation
uncorrected (shown in black) cases. We note that the correlation
coefficients are greater than 0.5 in time-dilation uncorrected
cases where BAT channel 1 is involved in the lag extraction. In
the time-dilation corrected case all plots show correlation
coefficients greater than 0.5 except for the Lag43 plot. Despite
these moderate correlation coefficients, the large scatter seen in
these plots indicate that the observer-frame lag does not directly
represent the source-frame lag.

\subsection{Lag-$L_{\rm iso}$ Relation: Observer-frame versus Source-frame}

There are two important changes in the Lag-Luminosity relation
which may occur when going from fixed observer-frame energy bands
to fixed source-frame energy bands: A change in the power-law
index, and a change in the dispersion of the data measured by the
correlation coefficient. Table~\ref{SlopeCorrTable} summarizes
these two parameters for various energy bands both in the
observer-frame and in the source-frame.


In the observer-frame the power-law index varies from $\sim 0.6$
to $\sim 1.8$, with mean around 1.3. In the source-frame the index
changes from 0.9 to 1.23 with a mean of $\sim$1.1. Meanwhile, the
correlation coefficient varies from 0.60 to 0.79 in the
observer-frame and in the source-frame it changes from 0.76 to
0.90. Hence, according to Table~\ref{SlopeCorrTable}, the
source-frame Lag-$L_{\rm iso}$ relation seems to be tighter than
the observer-frame case with a slope closer to one.

\begin{table*}
\centering
 \begin{minipage}{140mm}
  \caption{Observer-frame and source-frame slopes and correlation coefficients of the lag-$L_{\rm iso}$
  relation. Conservative 10\% uncertainty is assumed for cases without uncertainties. \label{SlopeCorrTable}}
  \begin{tabular}{@{}llcccl@{}}
  \hline
Energy Bands & Frame & Slope & Correlation Coefficient & Number of GRBs & Reference \\
 \hline
(0.3-1), (3-10) keV & Observer & 0.95$\pm$0.23 & - & 9 & \cite{Margutti2010}\\
(6-25), (50-400) keV & Observer & 1.16$\pm$0.07 & - 0.79$^{+0.16}_{-0.05}$ & 8 & \cite{Arimoto2010}\\
(15-25), (25-50) keV & Observer & 1.4$\pm$0.1 & - 0.63$\pm$0.06 & 21 & U10\\
(15-25), (50-100) keV & Observer & 1.5$\pm$0.1 & - 0.60$\pm$0.06 & 28 & U10\\
(15-25), (100-200) keV & Observer & 1.8$\pm$0.1 & - 0.67$\pm$0.07 & 27 & U10\\
(25-50), (50-100) keV & Observer & 1.2$\pm$0.1 & - 0.66$\pm$0.07 & 27 & U10\\
(25-50), (100-200) keV & Observer & 1.4$\pm$0.1 & - 0.75$\pm$0.07 & 25 & U10\\
(25-50), (100-300) keV & Observer & 1.14$\pm$0.1 &  - & 6 & \cite{norris2000}\\
(25-50), (100-300) keV & Observer & 0.62$\pm$0.04 &  -0.72$\pm$0.07 & 6 & \cite{Hakkila2008}\\
(50-100), (100-200) keV & Observer & 1.4$\pm$0.1 & - 0.77$\pm$0.08 & 22 & U10\\
(20-100), (100-500) keV & Source   & 1.23$\pm$0.07 & - 0.90$^{+0.12}_{-0.02}$ & 8 & \cite{Arimoto2010}\\
(100-200), (300-400) keV & Source   & 0.9$\pm$0.1 & - 0.76$\pm$0.06 & 22 & \cite{Ukwatta2010lag}\\
(100-150), (200-250) keV & Source   & 1.2$\pm$0.2 & - 0.82$\pm$0.05 & 24 & This work\\
\hline
\end{tabular}
\end{minipage}
\end{table*}

\subsection{Spectral Lag - $E_{\rm pk}$ Relation}

\begin{figure*}
\centering
\includegraphics[width=120mm]{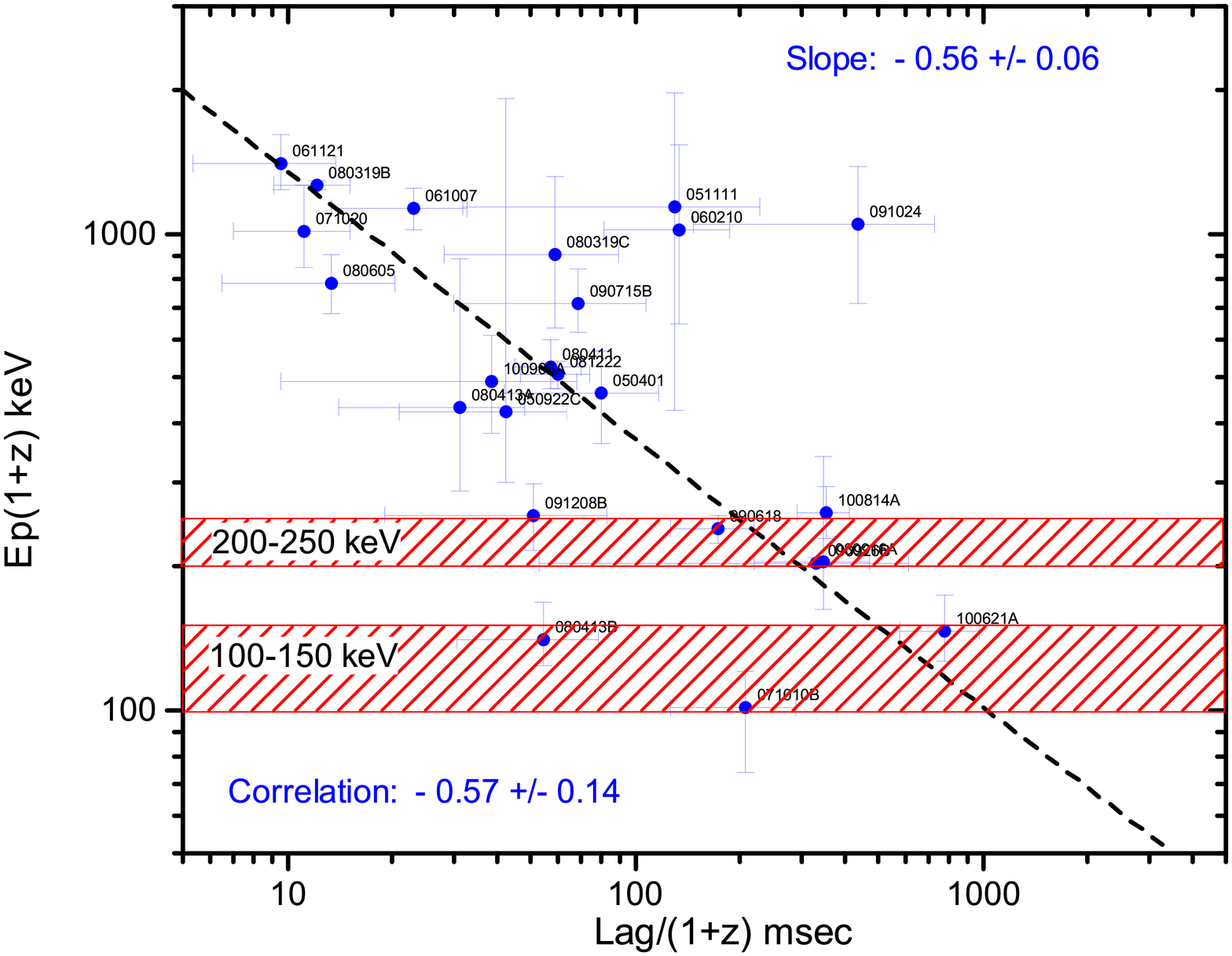}
\caption{The source-frame peak energy ($E_{\rm pk} (1+z)$) versus
source-frame spectral lags. The energy bands, $100-150$ keV and
$200-250$ keV, corresponding to the lag extractions are shown in
hashed red bands on the plot.}\label{EpSrcvsLagSrc}
\end{figure*}

Now we investigate the relation between source-frame spectral lag
and source-frame average peak energy ($E_{\rm pk}(1+z)$) of the
burst spectrum. In Fig.~\ref{EpSrcvsLagSrc}, we plotted $E_{\rm
pk}(1+z)$ as a function of source-frame lags. There is a
correlation between these two parameters with a correlation
coefficient of $-0.57 \pm 0.14$. Various correlation coefficients
of the relation are shown in Table~\ref{corr_table02}, with
uncertainties and null probabilities.

The best-fit is shown as a dashed line in
Fig.~\ref{EpSrcvsLagSrc}, yielding the following relation between
$E_{\rm pk}(1+z)$ and $Lag/(1+z)$:

\begin{equation}\label{eq:no3}
\log \bigg( \frac{E_{\rm pk}(1+z)}{\rm keV} \bigg) = (3.7 \pm 0.1)
- (0.56 \pm 0.06) \log \frac{ Lag/({\rm ms})}{1+z}.
\end{equation}

The uncertainties in the fitted parameters are expressed with the
factor of $\sqrt{\rm \chi^2/ndf} = \sqrt{30.71/22} \approx 1.18$.

According to equation~(\ref{eq:no2}), $L_{\rm iso} \propto
(Lag/(1+z))^{-1.2}$. From the Yonetoku relation we know that
$L_{\rm iso} \propto (E_{\rm
pk}(1+z))^{2.0}$~\citep{Yonetoku2004}. Hence, from these two
relations we expect to see a correlation between $E_{\rm pk}(1+z)$
and $Lag/(1+z)$ such as $E_{\rm pk}(1+z) \propto
(Lag/(1+z))^{-0.6}$.

The best fit slope of $0.56 \pm 0.06$ is consistent with the
expected slope of $\sim$ 0.6 based on the source-frame
lag-luminosity and the Yonetoku relation. However, note that the
correlation coefficient is significantly smaller than the
coefficient for the lag-luminosity relation. This lower degree of
correlation may be suggestive of brightness and detector related
selection effects that have been noted in the
literature~\citep{Butler2007} for the Yonetoku relation.

\begin{table}
\centering
 \begin{minipage}{84mm}
  \caption{Correlation coefficients of the lag--$E_{\rm pk}$ relation. \label{corr_table02}}
  \begin{tabular}{@{}lll@{}}
  \hline
  Coefficient Type     &  Correlation Coefficient & Null Probability \\
 \hline
Pearson's $r$          & - 0.57$\pm$0.14 & $4.83\,\times\,10^{-3}$\\
Spearman's $r_{\rm s}$ & - 0.50$\pm$0.12 & $1.36\,\times\,10^{-2}$\\
Kendall's $\tau$       & - 0.37$\pm$0.14 & $1.18\,\times\,10^{-2}$\\
\hline
\end{tabular}
\end{minipage}
\end{table}

\subsection{Some Models for Spectral Lags}

U10 and this work have provided more evidence for the existence of
the lag-luminosity relation based on a sample of $Swift$ BAT GRBs
with measured spectroscopic redshifts. This analysis calls for a
physical interpretation for spectral lag and a lag-luminosity
relation. In the literature, several possible interpretations have
been
discussed~\citep{Dermer1998,Salmonson2000,Ioka2001,Kocevski2003,Schaefer2004,Qin2004,Ryde2005,
Shen2005,Lu2006,Peng2011}.

\begin{figure*}
\centering
\includegraphics[width=90mm]{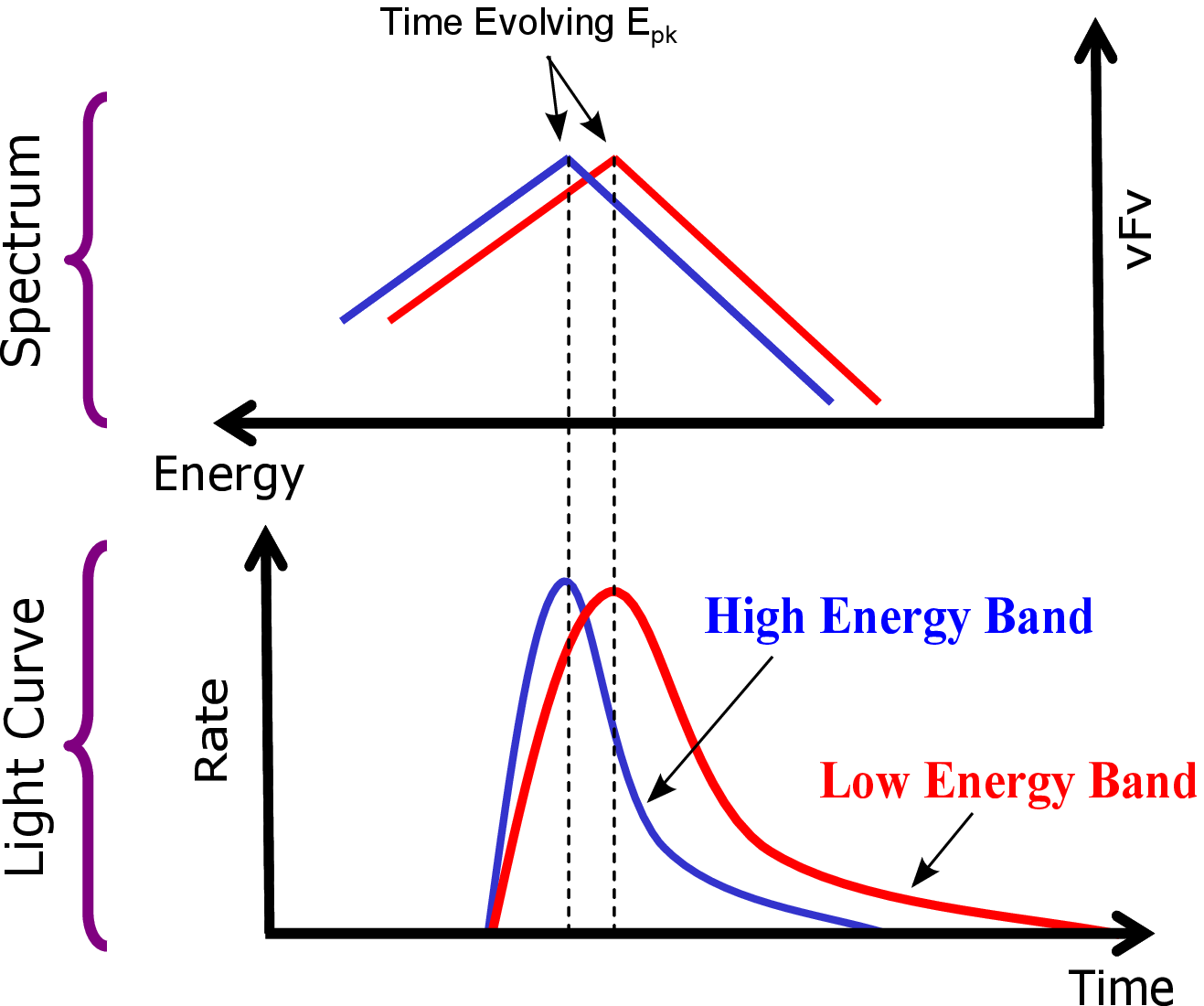}
\caption{The time evolution of the $E_{\rm pk}$ across energy
bands may cause the observed spectral lags in
GRBs.}\label{LagSpectralEvolution}
\end{figure*}

One proposed explanation for the observed spectral lag is the
spectral evolution during the prompt phase of the
GRB~\citep{Dermer1998, Kocevski2003, Ryde2005}. Due to cooling
effects, $E_{\rm pk}$ moves to a lower energy channel after some
characteristic time. When the peak energy ($E_{\rm pk}$) moves
from a higher energy band to a lower energy band, the temporal
peak of the light curve also moves from a higher energy band to a
lower one, which results in the observed spectral lag. In a recent
study, \cite{Peng2011} suggest that spectral evolution can be
invoked to explain both positive and negative spectral lags.
Hard-to-soft evolution of the spectrum produces positive spectral
lags while soft-to-hard evolution would lead to negative lags. In
addition, these authors also suggest that soft-to-hard-to-soft
evolution may produce negative lags.

A schematic diagram showing a hard-to-soft scenario is depicted in
Fig.~\ref{LagSpectralEvolution}. Initially, $E_{\rm pk}$ of the
spectrum is in the high-energy band, which results in a pulse in
the light curve of the high energy band. Then $E_{\rm pk}$ moves
to the lower energy band resulting in a pulse in the low-energy
light curve. The temporal difference between the two pulses in the
light curves would then be a measure of the cooling time scale of
the spectrum.

If this were the only process that caused the lag then in a simple
picture one would expect the source-frame average $E_{\rm pk}$ to
lie within the two energy bands in question. According to
Fig.~\ref{EpSrcvsLagSrc}, for the majority of bursts the
source-frame $E_{\rm pk}$ lies outside the energy band $100-250$
keV, indicating that the simple spectral evolution scenario
described above may not be the dominant process responsible for
the observed lags. However, it is worth noting that a pulse in a
specific energy band may not always mean that the $E_{\rm pk}$ is
also within that energy band. There are other issues associated
with this model: 1) the calculated cooling times based on simple
synchrotron models are, in general, relatively small compared to
the extracted lags, and 2) short bursts which exhibit considerable
spectral evolution do not show significant lags.

\begin{figure*}
\centering
\includegraphics[width=120mm]{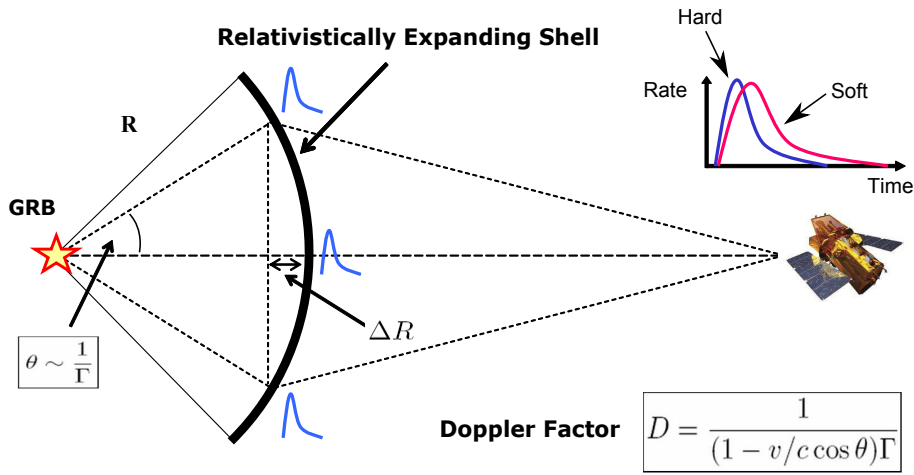}
\caption{Spectral lags could arise due to the curvature effect of
the shocked shell. At the source, the relativistically expanding
shell emits identical pulses from all latitudes. However, when the
photons reach the detector, on-axis photons get boosted to higher
energy (hard). Meanwhile off-axis photons get relatively smaller
boost and travel longer to reach the detector. Thus these photons
are softer and arrive later than the on-axis
photons.}\label{LagCurvatureEffect}
\end{figure*}

Another model that purports to explain spectral lags is based on
the curvature effect, i.e., a kinematics effect due to the
observer looking at increasingly off-axis annulus areas relative
to the
line-of-sight~\citep{Salmonson2000,Ioka2001,Dermer2004,Shen2005,Lu2006}.
Fig.~\ref{LagCurvatureEffect} illustrates how the spectral lag
could arise due to the curvature effect of the shocked shell. Due
to a smaller Doppler factor and a path difference, the radiation
from shell areas which are further off axis will be softer and
therefore lead to a lag. As with spectral evolution models, there
are difficulties associated with the curvature models too. These
kinematic models generally predict only positive lags. As can be
seen from Table~\ref{tab:lag} some of the measured lags are
negative, and therefore these lags present a real challenge for
the simple curvature models.

It is possible that spectral lags are caused by multiple
mechanisms. \cite{Peng2011} investigated spectral lags caused by
intrinsic spectral evolution and the curvature effect combined.
They showed that the curvature effect always tends to increase the
observed spectral lag in the positive direction. Even for cases
with soft-to-hard spectral evolution, when the curvature effect is
introduced lags become positive. Hence they predict that the
majority of measured spectral lags should be positive, which is
consistent with the findings of this work and U10.

\section[]{Summary and Conclusion} \label{conclusion}

We have investigated the spectral lag between $100-150$ keV and
$200-250$ keV energy bands at the GRB source-frame by projecting
these bands to the observer-frame. This is a step forward in the
investigation of lag-luminosity relations since most of the
previous investigations used arbitrary observer-frame energy
bands.

Our analysis has produced an improved correlation between spectral
lag ($\tau$) and isotropic luminosity over those previously
reported with the following relation:

\begin{equation}\label{conc:01}
\frac{\tau /(\rm ms)}{1+z} \cong \bigg[ \frac{L_{\rm iso}/(\rm
erg\,s^{-1})}{10^{54.7}}\bigg ]^{-0.8}.
\end{equation}

We also find a modest correlation between the source-frame
spectral lag and the peak energy of the burst, which is given by
the relation,

\begin{equation}\label{conc:02}
\frac{\tau /(\rm ms)}{1+z} \cong \bigg[ \frac{E_{\rm pk}(1+z)/(\rm
keV)}{10^{3.7}}\bigg ]^{-1.8}.
\end{equation}

Finally, we mentioned two simple models and noted their
limitations in explaining the observed spectral lags.

\section*{Acknowledgments}
We thank the anonymous referee for comments that significantly
improved the paper. The NSF grant 1002432 provided partial support
for the work of TNU and is gratefully acknowledged. The work of CD
is supported by the Office of Naval Research and Fermi Guest
Investigator grants. We acknowledge that this work has been
performed via the auspices of the GRB Temporal Analysis Consortium
(GTAC), which represents a comprehensive effort dedicated towards
the systematic study of spectral variation in Gamma-ray Bursts.


\begin{thebibliography}{99}
\bibitem[Arimoto et al.(2010)]{Arimoto2010} Arimoto, M., et al.\ 2010, \pasj, 62, 487
\bibitem[Band (1997)]{band1997} Band, D.~L.\ 1997, \apj, 486, 928
\bibitem[Band et al.(1993)]{band1993} Band, D., et al.\ 1993, \apj, 413, 281
\bibitem[Barthelmy et al.(2005)]{barthelmy2005} Barthelmy, S. D., et al. 2005a, Space Sci. Rev., 120, 143
\bibitem[Barthelmy et al.(2009)]{10103} Barthelmy, S.~D., et al.\ 2009, GRB Coordinates Network, Circular Service, 10103, 1
\bibitem[Barthelmy et al.(2010)]{11233} Barthelmy, S.~D., et al.\ 2010, GRB Coordinates Network, Circular Service, 11233, 1
\bibitem[Barthelmy et al.(2011)]{11714} Barthelmy, S.~D., et al.\ 2011, GRB Coordinates Network, Circular Service, 11714, 1
\bibitem[Baumgartner et al.(2009)]{10265} Baumgartner, W.~H., et al.\ 2009, GRB Coordinates Network, Circular Service, 10265, 1
\bibitem[Baumgartner et al.(2009)]{9775} Baumgartner, W.~H., et al.\ 2009, GRB Coordinates Network, 9775, 1
\bibitem[Baumgartner et al.(2009)]{9939} Baumgartner, W.~H., et al.\ 2009, GRB Coordinates Network, 9939, 1
\bibitem[Berger \& Becker(2005b)]{2005GCN..3520....1B} Berger, E., \& Becker, G.\ 2005, GRB Coordinates Network, 3520, 1
\bibitem[Briggs(2009)]{9957} Briggs, M.~S.\ 2009, GRB Coordinates Network, 9957, 1
\bibitem[Butler et al.(2007)]{Butler2007} Butler, N.~R., Kocevski, D., Bloom, J.~S., \& Curtis, J.~L.\ 2007, \apj, 671, 656
\bibitem[Cenko et al.(2007)]{Cenko2007} Cenko, S.~B., Cucchiara, A., Fox, D.~B., Berger, E., \& Price, P.~A.\ 2007, GRB Coordinates Network, 6888, 1
\bibitem[Cenko et al.(2009)]{2009GCN..9518....1S} Cenko, S. B., et al.\ 2009, GRB Coordinates Network, 9518, 1
\bibitem[Cenko et al.(2011)]{2011GCN.11638....1C} Cenko, S.~B., Hora, J.~L., \& Bloom, J.~S.\ 2011, GRB Coordinates Network, Circular Service, 11638, 1
\bibitem[Chaplin(2009)]{10095} Chaplin, V.\ 2009, GRB Coordinates Network, Circular Service, 10095, 1
\bibitem[Chen et al.(2009)]{2009GCN.10038....1C} Chen, H.-W., Helsby, J., Shectman, S., Thompson, I., \& Crane, J.\ 2009, GRB Coordinates Network, Circular Service, 10038, 1
\bibitem[Chornock et al.(2009)]{2009GCN..9243....1C} Chornock, R., Perley, D.~A., Cenko, S.~B., \& Bloom, J.~S.\ 2009, GRB Coordinates Network, 9243, 1
\bibitem[Chornock et al.(2009)]{2009GCN.10100....1C} Chornock, R., Perley, D.~A., \& Cobb, B.~E.\ 2009, GRB Coordinates Network, Circular Service, 10100, 1
\bibitem[Cucchiara \& Fox(2008)]{2008GCN..7654....1C} Cucchiara, A., \& Fox, D.~B.\ 2008, GRB Coordinates Network, 7654, 1
\bibitem[Cucchiara et al.(2006)]{2006GCN..4729....1C} Cucchiara, A., Fox, D.~B., \& Berger, E.\ 2006, GRB Coordinates Network, 4729, 1
\bibitem[Cucchiara et al.(2008)]{Cucchiara2008} Cucchiara, A., Fox, D.~B., Cenko, S.~B., \& Berger, E.\ 2008, GRB Coordinates Network, 8713, 1
\bibitem[Cucchiara et al.(2009)]{2009GCN.10065....1C} Cucchiara, A., Fox, D., \& Tanvir, N.\ 2009, GRB Coordinates Network, Circular Service, 10065, 1
\bibitem[D'Elia et al.(2009)]{DElia2009} D'Elia, V., et al.\ 2009, \apj, 694, 332
\bibitem[Dermer(1998)]{Dermer1998} Dermer, C.~D.\ 1998, \apjl, 501, L157
\bibitem[Dermer(2004)]{Dermer2004} Dermer, C.~D.\ 2004, \apj, 614, 284
\bibitem[Dermer \& Menon(2009)]{Dermer2009} Dermer, C.~D., \& Menon, G.\ 2009, High Energy Radiation from Black Holes: Gamma Rays, Cosmic Rays, and Neutrinos by Charles D.~Dermer and Govind Menon.~Princeton Univerisity Press, November 2009
\bibitem[Fenimore et al.(1995)]{Fenimore1995} Fenimore, E.~E., in 't Zand, J.~J.~M., Norris, J.~P., Bonnell, J.~T., \& Nemiroff, R.~J.\ 1995, \apjl, 448, L101
\bibitem[Foley(2011)]{11727} Foley, S.\ 2011, GRB Coordinates Network, Circular Service, 11727, 1
\bibitem[Fynbo et al.(2006)]{Fynbo2006} Fynbo, J.~P.~U., Limousin, M., Castro Cer{\'o}n, J.~M., Jensen, B.~L., \& Naranen, J.\ 2006, GRB Coordinates Network, 4692, 1
\bibitem[Fynbo et al.(2009)]{2009GCN..9947....1F} Fynbo, J.~P.~U., Malesani, D., Jakobsson, P., \& D'Elia, V.\ 2009, GRB Coordinates Network, 9947, 1
\bibitem[Fynbo et al.(2009)]{Fynbo2009} Fynbo, J.~P.~U., et al.\ 2009, arXiv:0907.3449
\bibitem[Gehrels et al.(2004)]{gehrels2004} Gehrels, N., et al. 2004, \apj, 611, 1005
\bibitem[Gehrels et al.(2006)]{gehrels2006} Gehrels, N., et al.\ 2006, \nat, 444, 1044
\bibitem[Golenetskii et al.(2009)]{10045} Golenetskii, S., Aptekar, R., Mazets, E., Pal'Shin, V., Frederiks, D., Oleynik, P., Ulanov, M., \& Svinkin, D.\ 2009, GRB Coordinates Network, Circular Service, 10045, 1
\bibitem[Golenetskii et al.(2009)]{10083} Golenetskii, S., et al.\ 2009, GRB Coordinates Network, Circular Service, 10083, 1
\bibitem[Golenetskii et al.(2010)]{10882} Golenetskii, S., et al.\ 2010, GRB Coordinates Network, Circular Service, 10882, 1
\bibitem[Golenetskii et al.(2010)]{11251} Golenetskii, S., et al.\ 2010, GRB Coordinates Network, Circular Service, 11251, 1
\bibitem[Golenetskii et al.(2011)]{11659} Golenetskii, S., et al.\ 2011, GRB Coordinates Network, Circular Service, 11659, 1
\bibitem[Hakkila et al.(2008)]{Hakkila2008} Hakkila, J., Giblin, T.~W., Norris, J.~P., Fragile, P.~C., \& Bonnell, J.~T.\ 2008, \apjl, 677, L81
\bibitem[Ioka \& Nakamura(2001)]{Ioka2001} Ioka, K., \& Nakamura, T.\ 2001, \apjl, 554, L163
\bibitem[Jakobsson et al.(2005)]{Jakobsson2005} Jakobsson, P., Fynbo, J.~P.~U., Paraficz, D., Telting, J., Jensen, B.~L., Hjorth, J., \& Castro Cer{\'o}n, J.~M.\ 2005, GRB Coordinates Network, 4029, 1
\bibitem[Jakobsson et al.(2007)]{2007GCN..6952....1J} Jakobsson, P., Vreeswijk, P.~M., Hjorth, J., Malesani, D., Fynbo, J.~P.~U., \& Thoene, C.~C.\ 2007, GRB Coordinates Network, 6952, 1
\bibitem[Jakobsson et al.(2008b)]{Jakobsson2008} Jakobsson, P., Vreeswijk, P.~M., Xu, D., \& Thoene, C.~C.\ 2008, GRB Coordinates Network, 7832, 1
\bibitem[Jaunsen et al.(2008)]{Jaunsen2008} Jaunsen, A.~O., et al.\ 2008, \apj, 681, 453
\bibitem[Kaneko et al.(2006)]{Kaneko2006} Kaneko, Y., Preece, R.~D., Briggs, M.~S., Paciesas, W.~S., Meegan, C.~A., \& Band, D.~L.\ 2006, \apjs, 166, 298
\bibitem[Kocevski \& Liang(2003)]{Kocevski2003} Kocevski, D., \& Liang, E.\ 2003, \apj, 594, 385
\bibitem[Komatsu et al.(2009)]{Komatsu2009} Komatsu, E., et al.\ 2009, \apjs, 180, 330
\bibitem[Kouveliotou et al.(1993)]{Kouveliotou1993} Kouveliotou, C., Meegan, C.~A., Fishman, G.~J., Bhat, N.~P., Briggs, M.~S., Koshut, T.~M., Paciesas, W.~S., \& Pendleton, G.~N.\ 1993, \apjl, 413, L101
\bibitem[Krimm et al.(2010)]{11094} Krimm, H.~A., et al.\ 2010, GRB Coordinates Network, Circular Service, 11094, 1
\bibitem[Lu et al.(2006)]{Lu2006} Lu, R.-J., Qin, Y.-P., Zhang, Z.-B., \& Yi, T.-F.\ 2006, \mnras, 367, 275
\bibitem[Margutti et al.(2010)]{Margutti2010} Margutti, R., Guidorzi, C., Chincarini, G., Bernardini, M.~G., Genet, F., Mao, J., \& Pasotti, F.\ 2010, \mnras, 406, 2149
\bibitem[Markwardt et al.(2009)]{10040} Markwardt, C.~B., et al.\ 2009, GRB Coordinates Network, Circular Service, 10040, 1
\bibitem[McBreen(2009)]{10266} McBreen, S.\ 2009, GRB Coordinates Network, Circular Service, 10266, 1
\bibitem[Milne \& Cenko(2011)]{2011GCN.11708....1M} Milne, P.~A., \& Cenko, S.~B.\ 2011, GRB Coordinates Network, Circular Service, 11708, 1
\bibitem[Milvang-Jensen et al.(2010)]{2010GCN.10876....1M} Milvang-Jensen, B., et al.\ 2010, GRB Coordinates Network, Circular Service, 10876, 1
\bibitem[Mosquera Cuesta et al.(2008)]{2008A&A...487...47M} Mosquera Cuesta, H.~J., Turcati, R., Furlanetto, C., Khachatryan, H.~G., Mirzoyan, S., \& Yegorian, G.\ 2008, \aap, 487, 47
\bibitem[Murakami et al.(2003)]{Murakami2003} Murakami, T., Yonetoku, D., Izawa, H., \& Ioka, K.\ 2003, \pasj, 55, L65
\bibitem[Norris (1995)]{norris1995} Norris, J.~P.\ 1995, \apss, 231, 95
\bibitem[Norris (2002)]{norris2002} Norris, J.~P.\ 2002, \apj, 579, 386
\bibitem[Norris \& Bonnell(2006)]{norris2006} Norris, J.~P., \& Bonnell, J.~T.\ 2006, \apj, 643, 266
\bibitem[Norris et al. (2000)]{norris2000} Norris, J.~P., Marani, G.~F., \& Bonnell, J.~T.\ 2000, \apj, 534, 248
\bibitem[Norris et al.(1996)]{norris1996} Norris, J.~P., Nemiroff, R.~J., Bonnell, J.~T., Scargle, J.~D., Kouveliotou, C., Paciesas, W.~S., Meegan, C.~A., \& Fishman, G.~J.\ 1996, \apj, 459, 393
\bibitem[Norris et al.(2005)]{norris2005} Norris, J.~P., Bonnell, J.~T., Kazanas, D., Scargle, J.~D., Hakkila, J., \& Giblin, T.~W.\ 2005, \apj, 627, 324
\bibitem[O'Meara et al.(2010)]{2010GCN.11089....1O} O'Meara, J., Chen, H.-W., \& Prochaska, J.~X.\ 2010, GRB Coordinates Network, Circular Service, 11089, 1
\bibitem[Pal'Shin et al.(2009)]{9821} Pal'Shin, V., et al.\ 2009, GRB Coordinates Network, 9821, 1
\bibitem[Palmer et al.(2009)]{10051} Palmer, D.~M., et al.\ 2009, GRB Coordinates Network, Circular Service, 10051, 1
\bibitem[Peng et al.(2011)]{Peng2011} Peng, Z.~Y., Yin, Y., Bi, X.~W., Bao, Y.~Y., \& Ma, L.\ 2011, Astronomische Nachrichten, 332, 92
\bibitem[Penprase et al.(2006)]{Penprase2006} Penprase, B.~E., et al.\ 2006, \apj, 646, 358
\bibitem[Piranomonte et al.(2008)]{Piranomonte2008} Piranomonte, S., et al.\ 2008, \aap, 492, 775
\bibitem[Prochaska et al.(2006)]{2006GCN..5002....1P} Prochaska, J.~X., Chen, H.-W., Bloom, J.~S., Falco, E., \& Dupree, A.~K.\ 2006, GRB Coordinates Network, 5002, 1
\bibitem[Prochaska et al.(2008)]{Prochaska2008} Prochaska, J.~X., Shiode, J., Bloom, J.~S., Perley, D.~A., Miller, A.~A., Starr, D., Kennedy, R., \& Brewer, J.\ 2008, GRB Coordinates Network, 7849, 1
\bibitem[Prochaska et al.(2009)]{Prochaska2009} Prochaska, J.~X., et al.\ 2009, \apjl, 691, L27
\bibitem[Qin et al.(2004)]{Qin2004} Qin, Y.-P., Zhang, Z.-B., Zhang, F.-W., \& Cui, X.-H.\ 2004, \apj, 617, 439
\bibitem[Rol et al.(2006)]{2006GCN..5555....1R} Rol, E., Jakobsson, P., Tanvir, N., \& Levan, A.\ 2006, GRB Coordinates Network, 5555, 1
\bibitem[Ryde(2005)]{Ryde2005} Ryde, F.\ 2005, \aap, 429, 869
\bibitem[Sakamoto et al.(2008)]{Taka2008} Sakamoto, T., et al.\ 2008, \apjs, 175, 179
\bibitem[Sakamoto et al.(2009)]{10072} Sakamoto, T., et al.\ 2009, GRB Coordinates Network, Circular Service, 10072, 1
\bibitem[Sakamoto et al.(2009)]{2009GCN..9231....1S} Sakamoto, T., et al.\ 2009, GRB Coordinates Network, 9231, 1
\bibitem[Sakamoto et al.(2009)]{Sakamoto2009} Sakamoto, T., et al.\ 2009, \apj, 693, 922
\bibitem[Sakamoto et al.(2011)]{Sakamoto2011} Sakamoto, T., et al.\ 2011, arXiv:1104.4689
\bibitem[Salmonson (2000)]{Salmonson2000} Salmonson, J.~D.\ 2000, \apjl, 544, L115
\bibitem[Salmonson \& Galama(2002)]{Salmonson2002} Salmonson, J.~D., \& Galama, T.~J.\ 2002, \apj, 569, 682
\bibitem[Sato et al.(2008)]{2008GCN..7591....1S} Sato, G., et al.\ 2008, GRB Coordinates Network, 7591, 1
\bibitem[Sbarufatti et al.(2008)]{2008GCNR..142.1S} Sbarufatti, B., et al.\ 2008, GCN Report, 142, 1
\bibitem[Schady et al.(2009)]{2009GCNR..232.1S} Schady, P., Baumgartner, W.~H., \& Beardmore, A.~P.\ 2009, GCN Report, 232, 1
\bibitem[Schaefer(2004)]{Schaefer2004} Schaefer, B.~E.\ 2004, \apj, 602, 306
\bibitem[Schaefer(2007)]{Schaefer2007} Schaefer, B.~E.\ 2007, \apj, 660, 16
\bibitem[Shen et al.(2005)]{Shen2005} Shen, R.-F., Song, L.-M., \& Li, Z.\ 2005, \mnras, 362, 59
\bibitem[Stamatikos et al.(2009)]{Stamatikos2009} Stamatikos, M., et al.\ 2009, American Institute of Physics Conference Series, 1133, 356
\bibitem[Tanvir et al.(2010)]{2010GCN.11230....1T} Tanvir, N.~R., Wiersema, K., \& Levan, A.~J.\ 2010, GRB Coordinates Network, Circular Service, 11230, 1
\bibitem[Ukwatta et al.(2010)]{291} Ukwatta, T.~N., et al.\ 2010, GCN Report, 291, 1
\bibitem[Ukwatta et al.(2010a)]{Ukwatta2009lag} Ukwatta, T.~N., et al.\ 2010a, \apj, 711, 1073 (U10)
\bibitem[Ukwatta et al.(2010b)]{Ukwatta2010lag} Ukwatta, T.~N., Dhuga, K.~S., Stamatikos, M., Sakamoto, T., Parke, W.~C., Barthelmy, S.~D., \& Gehrels, N.\ 2010b, arXiv:1003.0229
\bibitem[Watson et al.(2006)]{Watson2006} Watson, D., et al.\ 2006, \apj, 652, 1011
\bibitem[Wiersema et al.(2009)]{2009GCN..9673....1S} Wiersema, K., et al.\ 2009, GRB Coordinates Network, 9673, 1
\bibitem[Wiersema et al.(2009)]{2009GCN.10263....1W} Wiersema, K., Tanvir, N.~R., Cucchiara, A., Levan, A.~J., \& Fox, D.\ 2009, GRB Coordinates Network, Circular Service, 10263, 1
\bibitem[Xu et al.(2009)]{2009GCN.10053....1X} Xu, D., et al.\ 2009, GRB Coordinates Network, Circular Service, 10053, 1
\bibitem[Yonetoku et al.(2004)]{Yonetoku2004} Yonetoku, D., Murakami, T., Nakamura, T., Yamazaki, R., Inoue, A.~K., \& Ioka, K.\ 2004, \apj, 609, 935
\bibitem[Zhang et al.(2006)]{Zhang2006} Zhang, Z., Xie, G.~Z., Deng, J.~G., \& Jin, W.\ 2006, \mnras, 373, 729
\bibitem[Zhang et al.(2009)]{Zhang2009} Zhang, B., et al.\ 2009, \apj, 703, 1696
\bibitem[de Ugarte Postigo et al.(2009)]{2009GCN..9771....1D} de Ugarte Postigo, A., Gorosabel, J., Fynbo, J.~P.~U., Wiersema, K., \& Tanvir, N.\ 2009, GRB Coordinates Network, 9771, 1
\bibitem[von Kienlin(2010)]{11099} von Kienlin, A.\ 2010, GRB Coordinates Network, Circular Service, 11099, 1

\end{thebibliography}
\end{document}